\documentclass[twocolumn]{revtex4-2}
\usepackage{graphicx}% Include figure files
\usepackage{dcolumn}% Align table columns on decimal point
\usepackage{bm}% bold math
\usepackage{hyperref}% add hypertext capabilities
\usepackage{xcolor}
\usepackage{ulem}
\usepackage{cancel}
\usepackage{booktabs}
\hypersetup{
    colorlinks=true,
    linkcolor=blue,
    citecolor=blue,
    urlcolor=blue
}
\begin{document}

\preprint{APS/123-QED}
\title{\textbf{Temporal Correlation between Positive-Charged Cosmic Ray Flux and Solar Polar Field Variation: Insights from Delayed Modulation Analysis} 
}% 

\thanks{Corresponding author: Jie Feng, Zhibing Li}

% \email{E-mail: fengj77@mail.sysu.edu.cn, stslzb@mail.sysu.edu.cn}

\author{Shaokun Gong}
\affiliation{ School of Science, Shenzhen Campus of Sun Yat-sen University, Shenzhen 518107, China}

\author{Linjing Duan}
\affiliation{ School of Science, Shenzhen Campus of Sun Yat-sen University, Shenzhen 518107, China}
\author{Jiawei Zhao}
\affiliation{ School of Science, Shenzhen Campus of Sun Yat-sen University, Shenzhen 518107, China}
\author{Xueyu Wei}
\affiliation{ School of Science, Shenzhen Campus of Sun Yat-sen University, Shenzhen 518107, China}

\author{Jie Feng}
\email{fengj77@mail.sysu.edu.cn}
\affiliation{ School of Science, Shenzhen Campus of Sun Yat-sen University, Shenzhen 518107, China}

\author{Zhibing Li}
\email{stslzb@mail.sysu.edu.cn}
\affiliation{ School of Science, Shenzhen Campus of Sun Yat-sen University, Shenzhen 518107, China}

\date{\today}% It is always \today, today,

\begin{abstract}
We present an analysis of the time-dependent modulation of galactic cosmic rays near Earth, with a focus on the cosmic proton flux and polar field. Using data from the Alpha Magnetic Spectrometer (AMS) and the Wilcox Solar Observatory, we identify a significant time-lagged relationship between the observation of two missions. Our model incorporates a weighted magnetic field parameter to address the hemispheric asymmetry in polar fields and captures the temporal evolution of cosmic-ray proton spectra in relation to solar activity. We find a time lag of approximately 10 months, varying with cosmic ray rigidity.  At 1 GV, the time lag is 360 days, while it is 300 days above 3 GV. A potential mechanism is proposed to explain the observed time-lagged relationship and its dependence on cosmic ray rigidity.
This offers predictive insights into cosmic ray modulation within the heliosphere. These results enhance the accuracy of space weather forecasting models, with significant implications for the safety of space missions and aviation. 

\end{abstract}

\maketitle

%\tableofcontents
\section{Introduction} \label{sec:intro}
In recent years, the precision of cosmic-ray (CR) detection experiments has significantly improved, and measuring temporal variations in galactic cosmic rays has become increasingly important. These variations provide crucial insights into the dynamic processes within the heliosphere. Solar activities, such as solar wind, sunspot numbers, changes in the polar field, and solar modulation effects, significantly impact the energy spectra of CR. Therefore, studying these changes is essential for understanding CR propagation in the heliosphere.

The necessity of measuring temporal variations in low-energy galactic cosmic rays lies in their correlation with the solar activity cycle, particularly the quasi-periodic 11-year solar cycle, which affects the flux and energy distribution of galactic cosmic rays reaching the Earth. Observationally, the relationship between solar activity and cosmic-ray flux intensity has been widely validated. This relationship underscores the importance of continuous and precise CR measurements to understand the underlying physical mechanisms.

As galactic cosmic rays enter the heliosphere, time-dependent structures in their energy spectra are expected to arise due to solar modulation. This modulation involves several physical processes, including convection, diffusion, particle drift, and adiabatic energy changes. The complex interplay of these processes alters the energy spectra of galactic cosmic rays (GCRs) as they propagate through the heliosphere, making it essential to study solar modulation in order to understand the observed variations in cosmic-ray fluxes \citep{Tomassetti2015, 2015ApJ...803L..15T, PhysRevD.94.123007}.

Globally, several experiments and missions have contributed to this field by providing extensive data on galactic cosmic rays. For instance, Voyager 1 \citep{Webber18, 2014ApJ...794...29P} was the first to measure galactic cosmic rays in interstellar space, while long-term missions like Payload for Antimatter Matter Exploration and Light-nuclei Astrophysics (PAMELA) \citep{Martucci:2018pau, Adriani_2013} and AMS \citep{AGUILAR20211} have been continuously monitoring galactic cosmic rays. Additionally, EPHIN on the Solar and Heliospheric Observatory (SOHO) \citep{Kühl2016} has provided invaluable insights into solar and heliospheric conditions that influence cosmic-ray propagation. These missions have significantly enriched our understanding by offering time-resolved data on cosmic-ray particles and antiparticles. Ground-based observations also play a vital role, with neutron monitors and other detectors providing continuous data that complement space-based observations. The comprehensive datasets collected by these observatories have enabled the development of detailed models of CR propagation and solar modulation \citep{POTGIETER20141415}, enhancing our understanding of heliospheric processes \citep{2004ApJ...603..744F}.

Building on these extensive datasets, recent studies have further explored the temporal relationship between solar activity and cosmic-ray modulation. Refs.~\citep{tomassetti2017evidence, PhysRevD.106.103022} provided empirical evidence of a time delay between cosmic-ray modulation and solar activity indices, attributing this delay to heliospheric modulation effects. These foundational works highlighted the time-dependent nature of cosmic-ray modulation and pointed to the need for accurately modeling the delay in solar activity's influence on cosmic-ray flux. Expanding on this, Ref.~\cite{Koldobskiy2022} analyzed the time lags between monthly cosmic-ray intensity and various solar indices, including sunspot numbers and open solar flux, using cross-correlation and wavelet coherence methods. Their study revealed a significant delay in the cosmic-ray response to polar activity, with variations depending on solar polarity and heliospheric conditions. Together, these studies have laid crucial groundwork for understanding the timing effects in cosmic-ray modulation.

In this letter, utilizing an extensive dataset of modulated galactic cosmic-ray measurements collected from the Alpha Magnetic Spectrometer, we have discovered a significant correlation between solar activity, particularly the polar field, and the galactic cosmic rays. This correlation can aid in the development of predictive models for GCR modulation. Our approach incorporates a "delayed" relationship to capture the impact of solar activity observations and the conditions within the modulation region, estimating a time lag $\Delta T$ of approximately one year. Previous empirical studies have highlighted the significance of this lag in cosmic-ray modulation models \citep{tomassetti2017evidence}. We will illustrate that recent direct measurements of galactic cosmic rays indicate a lag of about ten months relative to solar magnetic activity. This finding establishes the timescale for the evolving conditions in the heliosphere, allowing us to predict near-Earth cosmic-ray fluxes with considerable lead time.

Our findings are crucial for the development of models to predict space weather effects, which are becoming a significant concern for both space missions and air travelers. These results not only contribute to scientific knowledge in plasma and solar astrophysics but also address practical issues related to the safety and reliability of space and aviation operations.

This letter is organized as the following. 
The analytical approach to the correlation between the polar field and the proton fluxes is presented in Section \ref{sec:methodology}.
The delayed effect of the polar field on the proton flux is presented in Section \ref{sec:results}.
We summarize our study in Section \ref{sec:conclusion}.
\section{Methodology} \label{sec:methodology}
Numerous studies have established that the polar field significantly influences the heliospheric magnetic field (HMF), thereby affecting the modulation of CRs entering the heliosphere. The HMF modulates CRs through various processes \citep{2005JGRA..11012108U, 2013LRSP...10....3P}, including drift, diffusion, convection, and adiabatic energy loss, altering their energy spectra and spatial distribution \citep{2004ApJ...603..744F, https://doi.org/10.1029/98GL00499}.

Ref.\cite{Putri2024} discusses the role of the polar field polarity in galactic cosmic ray (GCR) modulation. The solar cycle alternates between two distinct phases based on the polarity of the polar field: positive polarity ($A > 0$) and negative polarity ($A < 0$). These polarity changes significantly influence the heliospheric magnetic field (HMF) structure, affecting the drift patterns of charged particles. During positive polarity periods, protons of galactic cosmic rays enter the heliosphere through the poles, while during negative polarity periods, they drift along the heliospheric current sheet.\citep{jokipii1977effects, 2013LRSP...10....3P}. Therefore, the modulation effects of galactic cosmic rays must be analyzed separately for positive and negative polarity periods.

An apparent asymmetry in the effects of the northern and southern solar pole magnetic fields is observed, primarily due to seasonal variations in the measurements caused by Earth’s position relative to the Sun. The total effect of the polar field during solar minimum should not show seasonal variation, so this observed asymmetry is an observational effect. These variations must be accounted for to accurately interpret the data and model their impact on Galactic cosmic rays.

To address this asymmetry, we introduce a weighting parameter $w$ to adjust the influence between the magnetic field of the northern hemisphere ($B_N$) and the southern hemisphere ($B_S$). The weighted magnetic field $B_{\text{weighted}}$ is defined as:

\begin{equation}
B_{\text{weighted}} = \frac{B_N - wB_S}{1+w},
\label{beff}
\end{equation}
As \( w \rightarrow 0 \), the magnetic field of the northern hemisphere ($B_N$) predominates. Conversely, as \( w \rightarrow \infty \), the magnetic field of the southern hemisphere  ($B_S$) becomes dominant. When $ w = 1 $, the effects on the Galactic cosmic ray from both hemispheres are equal.

The polar field data utilized in this research was sourced from the publicly accessible archives of the Wilcox Solar Observatory \citep{Scherrer1977}. Data points were systematically recorded every 10 days.

\begin{figure}[ht!]
    \centering
    \includegraphics[width=0.45\textwidth]{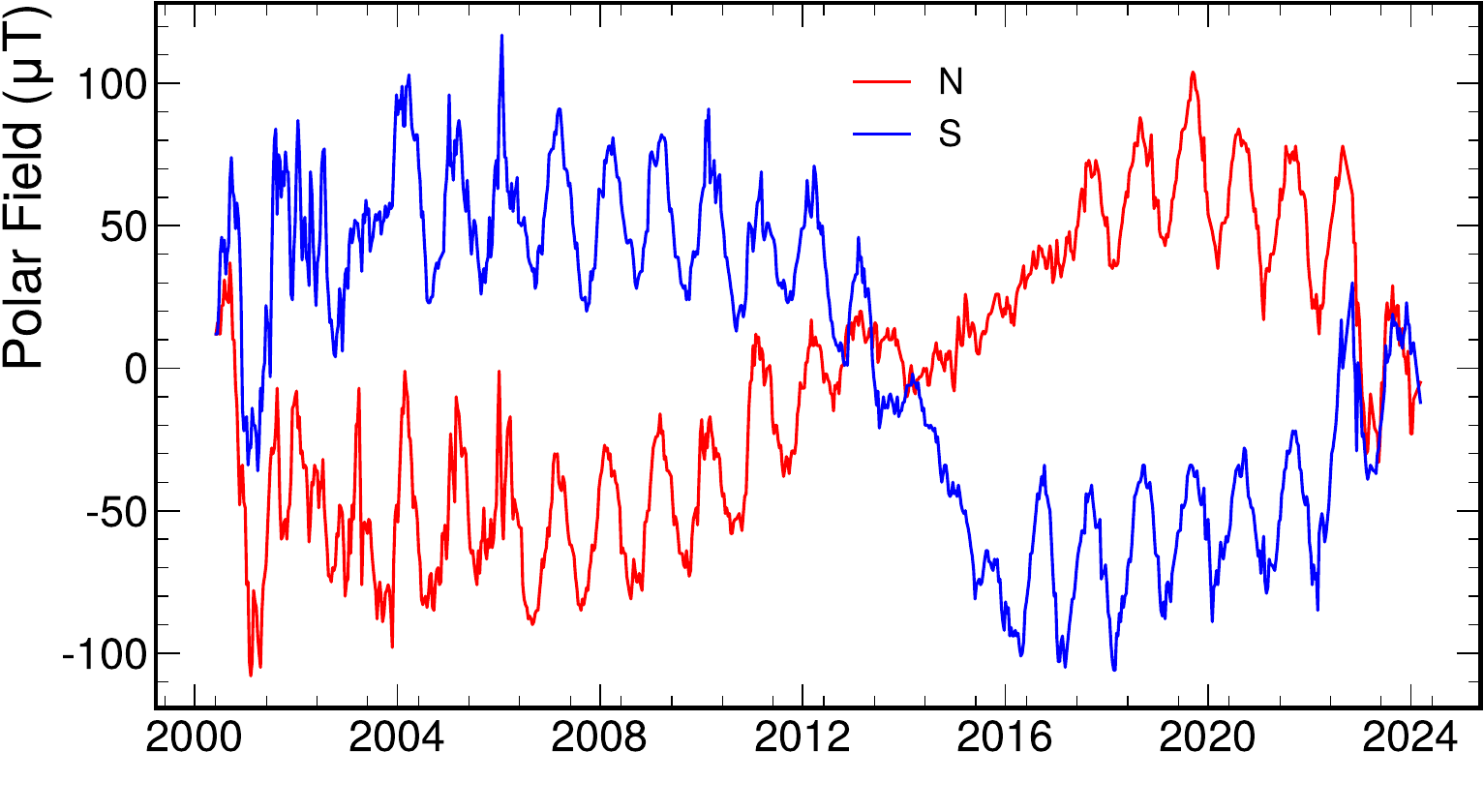} 
    \caption{The Wilcox Solar Observatory \citep{Scherrer1977} records the variations in the polar field \sout{strength} at the solar north and south poles from 2000 to 2024.}
    \label{fig:1}
\end{figure}

During the period from 2017 to 2020, the magnetic field distribution remained relatively stable, with no significant trends, as shown in Fig.\ref{fig:1}. To refine our analysis and mitigate variability, we adjusted $w$ to smooth the magnetic field data, reducing noise and non-periodic variations in Galactic cosmic ray flux. This adjustment enhances model accuracy by better capturing the modulation effects of the polar field. Notably, the proton flux during this period showed no large time structure in 10-day bins. We began with $w=1$, assuming equal weighting of the northern and southern magnetic fields, followed by a linear fit and a calculation of the  $\chi^2$.

Next, we varied the weight $ w $ within the range of 0.5 to 1.2. We re-fitted the magnetic field data with a linear function for each value of $ w $ and computed the differences in  $\chi^2$ for each weight. This iterative approach allowed us to identify the optimal weighting parameter  $w$  that minimized the variability and provided the smoothest temporal distribution of the polar fields.
The variance $\chi^2$ is calculated using the following formula:

\begin{equation}
   \chi^2(w) = \sum_{i} \left( \frac{y_i(w) - y_{\text{fit}, i}}{\sigma} \right)^2,
\label{weightchi}
\end{equation}
where $y_i(w)$ is the effective magnetic field, $y_{\text{fit}, i} $ is the value from the linear fit, and $i$ loops from 2017 to 2020, including 111 data points. $\sigma$ represents the uncertainty of the polar field measurements, which is taken as the standard deviation of the mean polar field from 2017 to 2020 when the magnetic field distribution is assumed to be relatively stable during solar minima. We obtain that $\sigma=4.18\,\mu $T, consistent with the previous estimations, which suggested it to be less than $5\,\mu $T. \cite{Ye2012, Boberg2002}

To ensure consistency in our analysis of the polar field’s influence on Galactic cosmic rays, we addressed the polarity reversal that occurred around 2013. Observational data and theoretical models indicate that the polar field underwent a polarity reversal during this period, switching from negative to positive polarity \citep{Mordvinov2014, Pishkalo2019, Aslam_2023}. We defined the post-2013 magnetic field as having a positive polarity. To maintain a consistent polarity for the entire dataset, we took the absolute value of the pre-2013 magnetic field measurements, effectively normalizing the data to a uniform polarity standard. Additionally, the weights $w$ were set differently for the periods before and after the polarity reversal, reflecting the changes in the magnetic field’s influence.

Furthermore, to align the frequency with the polar field data, 
the 10-day data are evaluated with a weighted average method with the weight equal to the inverse of the square of sigma error\citep{Ebert2022}. Converting it into ten-day average flux data.  Fig.~\ref{fig:combined2} shows that the time distribution of proton flux exhibits a very similar pattern to the time distribution of the polar field. We can define a time lag function as follows. Let Y denote the proton flux and X denote the weighted polar field determined by Eq.\ref{beff}. We introduce the following equation to model the relationship:

\begin{equation}
    Y(t)=aX(t-\Delta T)+b
\end{equation}
Here, $\Delta T$ represents the time lag between the solar activity indices and the medium properties of the modulation. This function allows us to quantify the delay between changes in polar activity and corresponding responses in galactic cosmic ray flux.

Our model is defined by three free parameters: $a,b,\Delta T$, which are constrained using extensive data. This data includes proton measurements collected by the AMS experiment from May 2011 to December 2019. At a given rigidity bin, each data point $J(t_j)$ represents the cosmic-ray flux at a specific time $t_j$, while $\hat{J}_{j}$ denotes the predicted value from the linear fit. The calculations incorporate delayed functions of the physical inputs. At this rigidity, we utilize a global $\chi^2$ estimator to determine the optimal time delay.
\begin{equation}
    \chi^{2}(a, b, \Delta T)=\sum_{j}\left[\frac{J\left(t_{j} ; a, b, \Delta T\right)-\hat{J}_{j}}{\sigma_{j}}\right]^{2},
\end{equation}
The total uncertainty in the flux, denoted by $\sigma_j$, is typically the quadrature sum of the statistical uncertainty and the time-dependent systematic uncertainty.
By adjusting $\Delta T$ to find the value that minimizes $\chi^2$, we determine the time delay relationship between solar magnetic activity and galactic cosmic rays, 

According to theoretical references, the error range of the $\chi^2$ distribution is determined when the $\chi^2$ increases a certain specified value \citep{Behnke2013, Gregory2005}. For a given set of data, the confidence region is defined by the inequality:

\begin{equation}
\chi^2(\theta) \leq \chi^2_{\text{min}} + \Delta \chi^2.
\label{error}
\end{equation}

Here, $\Delta \chi^2$ is typically derived from the $\chi^2$ distribution based on the desired confidence level and the number of degrees of freedom in the model. 

This ensures that the error range corresponds to the interval in which the parameter values are statistically likely to be found, with the observed data on hand.

To analyze the rigidity and temporal dependence of the modulation lag, we modify the global formula proposed in \citep{PhysRevD.106.103022}, as shown below:
\begin{equation}
{\Delta T} = {\Delta T}_{\text{Min}} + {\Delta T}_{\text{M}} \left(\frac{R}{\text{GV}}\right)^{-\delta}
\label{fit}
\end{equation}
This formula captures the combined effects of rigidity and temporal evolution on the modulation lag.
    
We used proton data spanning 2000 days combined with polar field data from 2014 to 2019.  The weighted average method was employed to calculate the mean flux of 10 days. To find the optimal time shift that minimizes the \(\chi^2\) value, we shifted the proton flux data backward and forward by a total of 730 days. Each shift corresponds to one data point, which is equivalent to 10 days, resulting in 73 shifts in total.

\section{Results} \label{sec:results}
Firstly, we calculated the weight of the polar field. Using relatively stable data points from 2017 to 2020, 111 points, we performed a linear fit to obtain the results. Without applying any weight, the composite magnetic field had a $\chi^2/ndf= 108.06/108$.

\begin{figure}[h]
\centering
\includegraphics[width=1\linewidth]{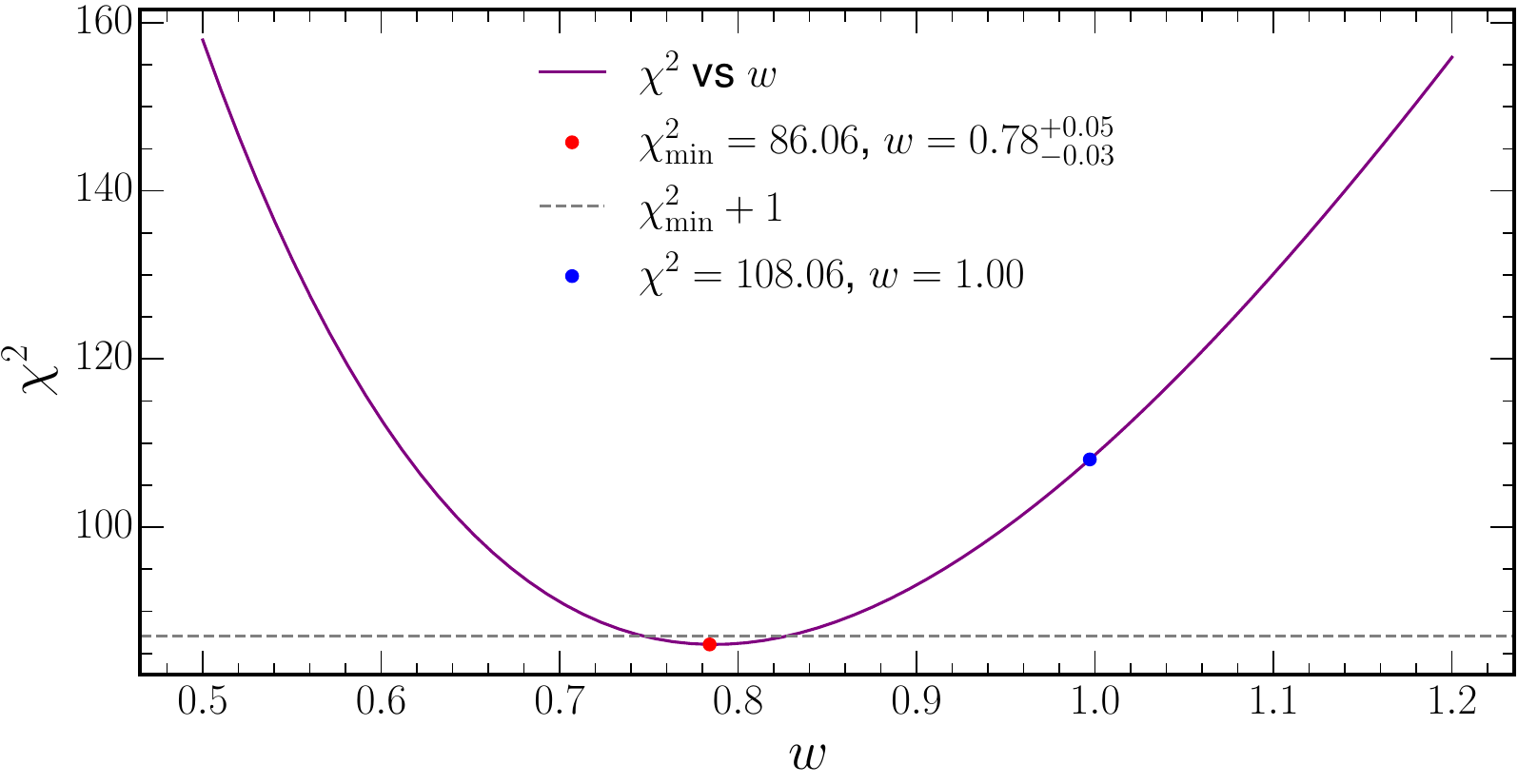}
\caption{$\chi^2$ distribution as a function of the effective weight of the magnetic fields. The red point shows the minimized $\chi^2$, while the blue point shows the $\chi^2$ with $w = 1.00$, where the solar north and south magnetic fields have the same weight.
\label{fig:generalll}}
\end{figure}

In Fig.\ref{fig:generalll}, by using Eq.\ref{weightchi}  to find the minimum $\chi^2/ndf$, we achieved a value of $86.06/108$, corresponding to an optimal weight of $w=0.78^{+0.05}_{-0.03}$. The difference in the $\chi^2$ compared to the unoptimized case $(w = 1.00)$ yields a highly significant p-value of $2.73 \times 10^{-6}$. This indicates that the optimization substantially improves the fit of the model. By setting this weight, we smoothed the composite total polar field, thereby eliminating fluctuations irrelevant to galactic cosmic ray modulation.
The difference in chi-square values around 6 is significant, indicating a clear discrepancy.

After determining the optimal weight for the $A>0$ period, the global fit has been performed on 2560 proton data points collected between 2014 and 2019 (in $A > 0$ conditions) at rigidity intervals between 1 and 33.5 GV \citep{PhysRevLett.127.271102}. The best-fit parameters for each of the 27 rigidity intervals were determined. The optimal fit parameters in the 1.00-1.16 GV interval are  $a = 19.3 \pm 0.1,  b = 349 \pm 2$, and  $\chi^2/ndf = 3621.0/198 $, with a time delay of  $360 \pm 5 $ days as determined by the confidence region by using Eq.\ref{error}. When the time delay was not set, the fit results were $ a = 17.2 \pm 0.1,  b = 408 \pm 3 $, and  $\chi^2/ndf = 43116.3/198$. These results are based on the positive polarity period data from 2014 to 2019.

\begin{figure}[ht!]
    \centering
 %   \begin{subfigure}{\linewidth}
 
        \includegraphics[width=0.9\linewidth]{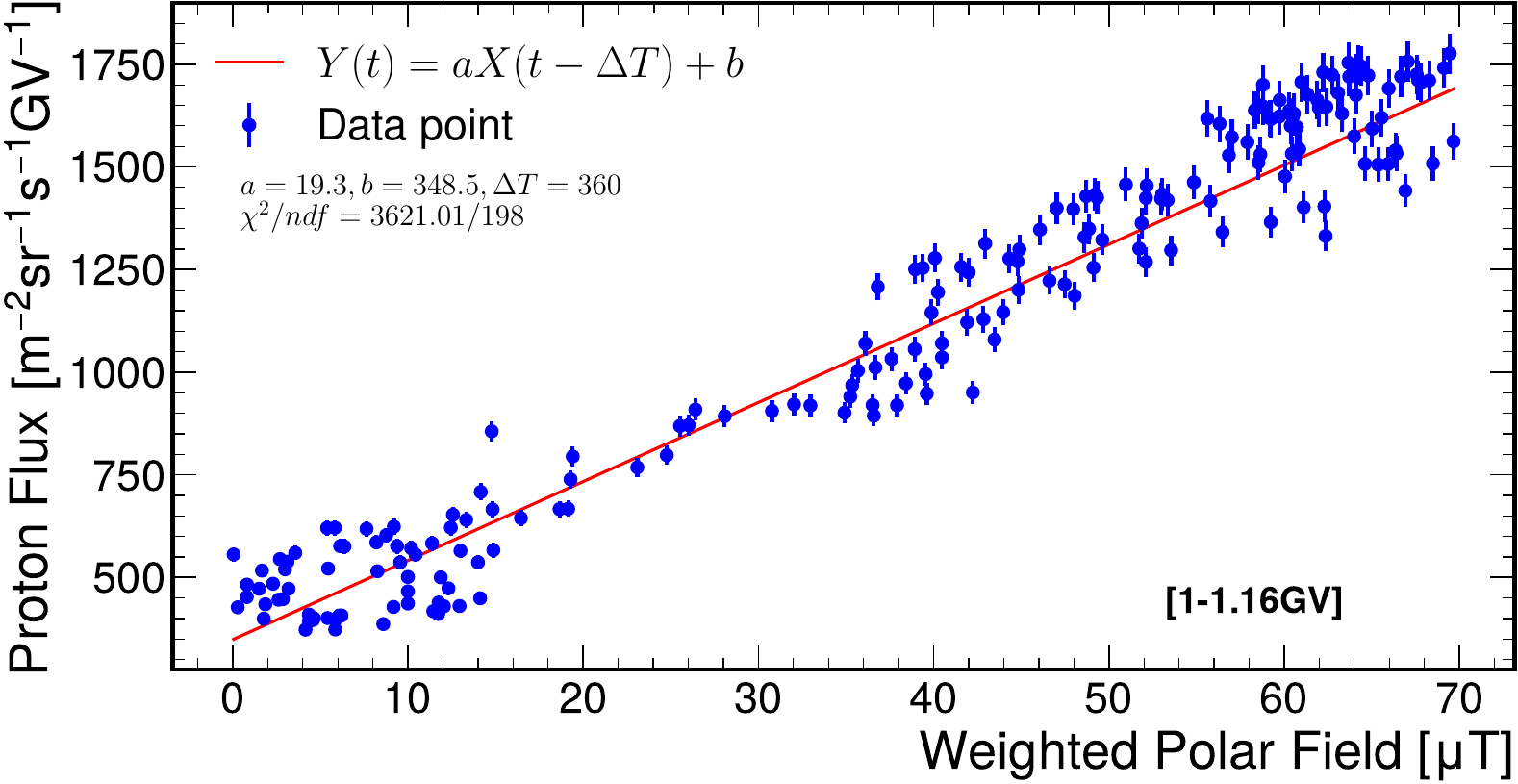}
        \includegraphics[width=0.9\linewidth]{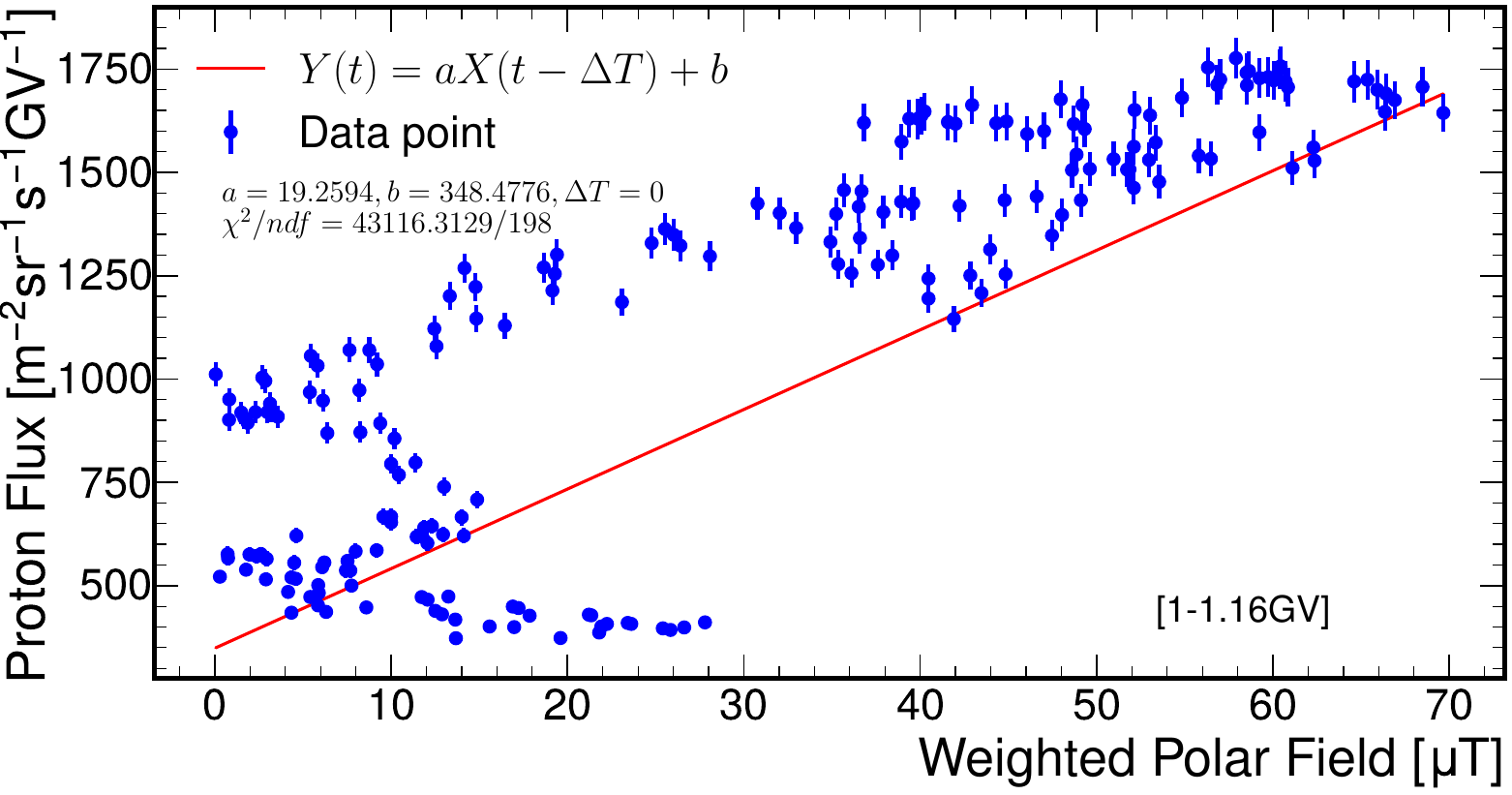}

    \caption{Comparison of the weighted polar field versus the proton flux in the rigidity range 1.00-1.16 GV. The top panel shows the original ($\Delta T = 0$) distribution compared with a linear function ($\chi^2/ndf=43116.3/198$), while the bottom panel shows the shifted ($\Delta T = 360 $ days) distribution ($\chi^2/ndf=3621.0/198$).}
    \label{fig:combined1}
 \end{figure}

In Fig.\ref{fig:combined1}, the optimal time lag adjustment results in a significantly better fit to the data, reducing the $\chi^2$ value from 43116 to 3621. This substantial reduction in $\chi^2$ value indicates that the time lag plays a crucial role in accurately modeling the relationship between the polar field and galactic cosmic ray proton flux, resulting in a better linear relationship.

\begin{figure}[ht!]
    \centering
        \includegraphics[width=0.9\linewidth]{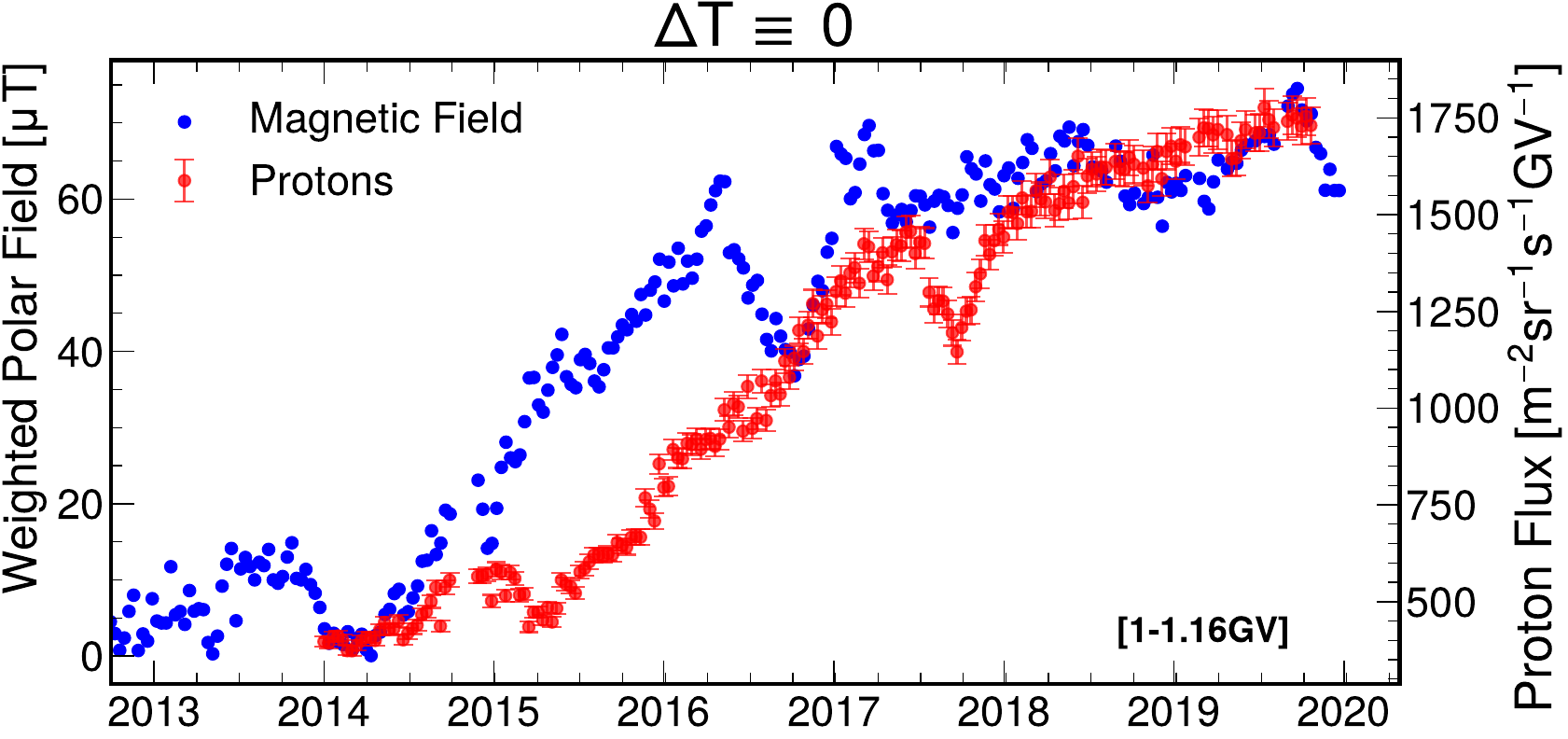}
        \includegraphics[width=0.9\linewidth]{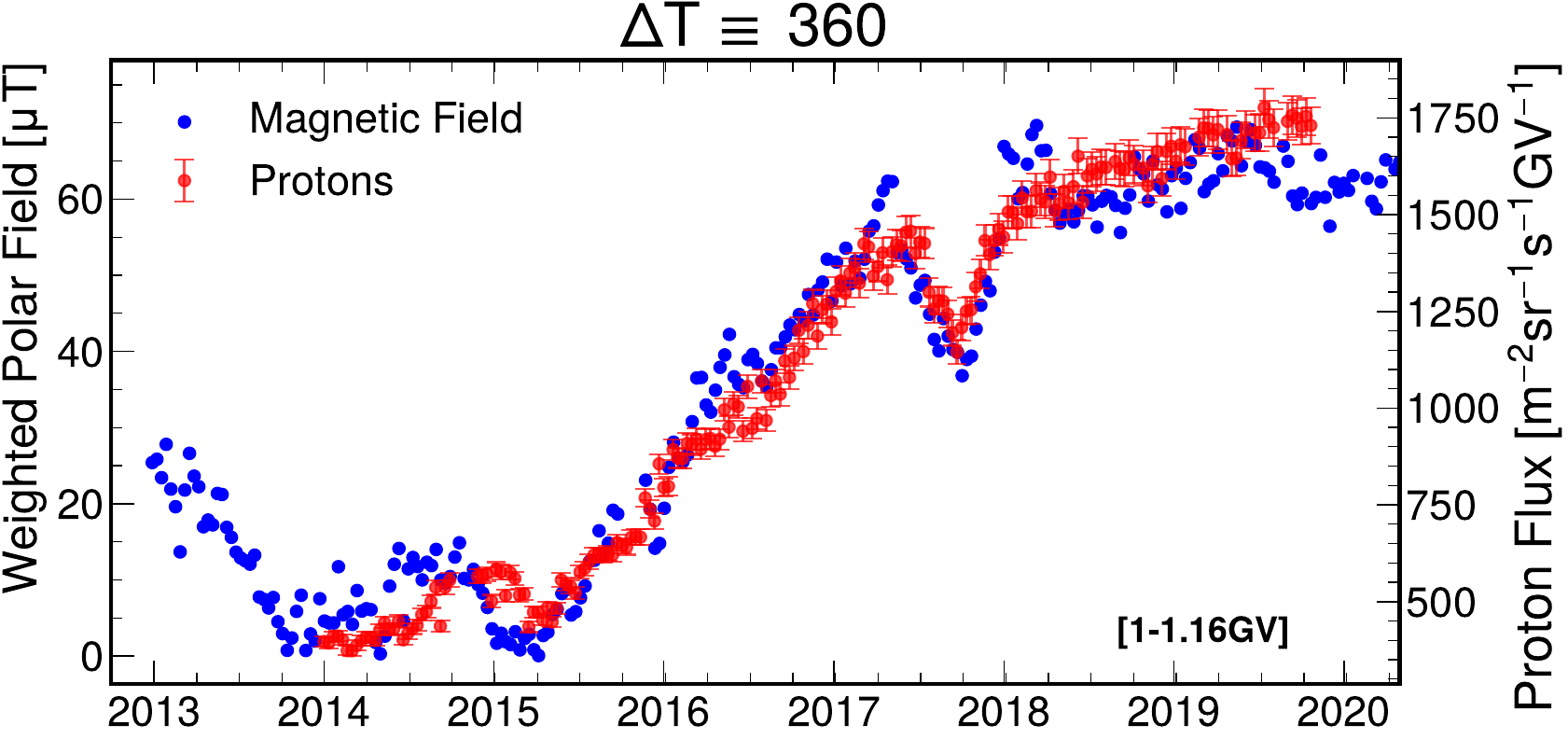}
    \caption{Comparison of GCR proton flux and weighted polar field for rigidity 1.00-1.16 GV. The blue points represent the polar field, and the red points represent the flux of protons.  The top panel shows the polar field and the proton fluxes do not vary simultaneously, while the bottom panel shows that they vary in the same trend if a delay of 360 days is applied to the magnetic field.}
    \label{fig:combined2}
\end{figure}

Fig.\ref{fig:combined2} further supports this finding by showing the time series comparison between the weighted polar field and GCR proton flux. The top panel presents the relationship without accounting for the time lag, while the bottom panel displays the relationship with the optimal time lag applied. The blue points represent the polar field, and the red points represent the flux of protons. With the optimal time lag, the alignment between the variations in the magnetic field and proton flux is significantly improved.

The relationship between the weighted polar field and galactic cosmic ray flux was analyzed using absolute values of the magnetic field, ensuring that only the field’s magnitude influences the results.

\begin{figure}[ht]
\centering
\includegraphics[width=0.75\linewidth]{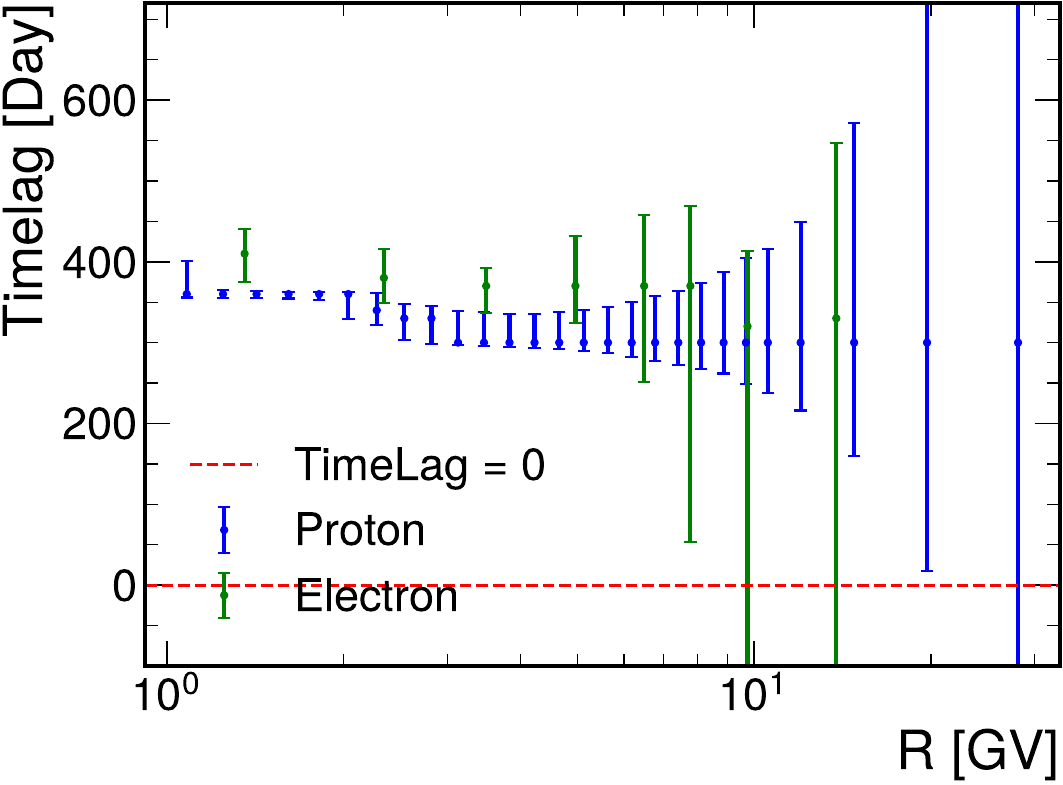}
\caption{Rigidity dependence of the time lag. It includes the error determined using Eq.\ref{error}. The points indicate the time delay length where $\chi^2$ is minimized. The red dash line along 0 is to guide your eye.
\label{fig:general}}
\end{figure}

We performed multiple calculations in different rigidity intervals and observed a trend where the time delay decreases with increasing rigidity \citep{PhysRevD.106.103022}. This effect is more pronounced in the 1-5 GV interval, as shown in Fig.\ref{fig:general}. As rigidity increases, the time delay tends to stabilize. However, due to the increasing measurement errors of AMS with higher rigidity, the errors defined by the confidence region in Eq.\ref{error} also increase. Specifically, in the 1.00-1.16 GV rigidity interval, the optimal time delay is $360 ^{+40}_{-2}$ days. In the 13-16.6 GV interval, the optimal delay is  $300^{+272}_{-141}$ days.

Additionally, we analyzed the correlation between the polar field and electron fluxes for the same period, as shown in Fig.\ref{fig:general}. For rigidity ranges of 1–1.71 GV and 7.09-8.48 GV, the calculated time delays are $410^{+31}_{-35}$ days and $370^{+99}_{-317}$ days respectively.

In this study, we focused on the time delay relationship between galactic cosmic rays and the polar field during the $A>0$ period (2014-2020). Our calculations indicate a time delay ranging from 360 days to 300 days, depending on rigidity. At rigidity levels above 22.8 GV, the time lag is no longer significant. During the polar field reversal period \citep{2013LRSP...10....5O}, the constantly changing polarity makes it challenging to establish a fixed time delay pattern to explain the discrepancies between galactic cosmic rays and the polar field \citep{ZERBO2013265, 04f6fe1bdcd94df68916164f66e6e018}. 

\begin{figure}[ht]
\centering
\includegraphics[width=\linewidth]{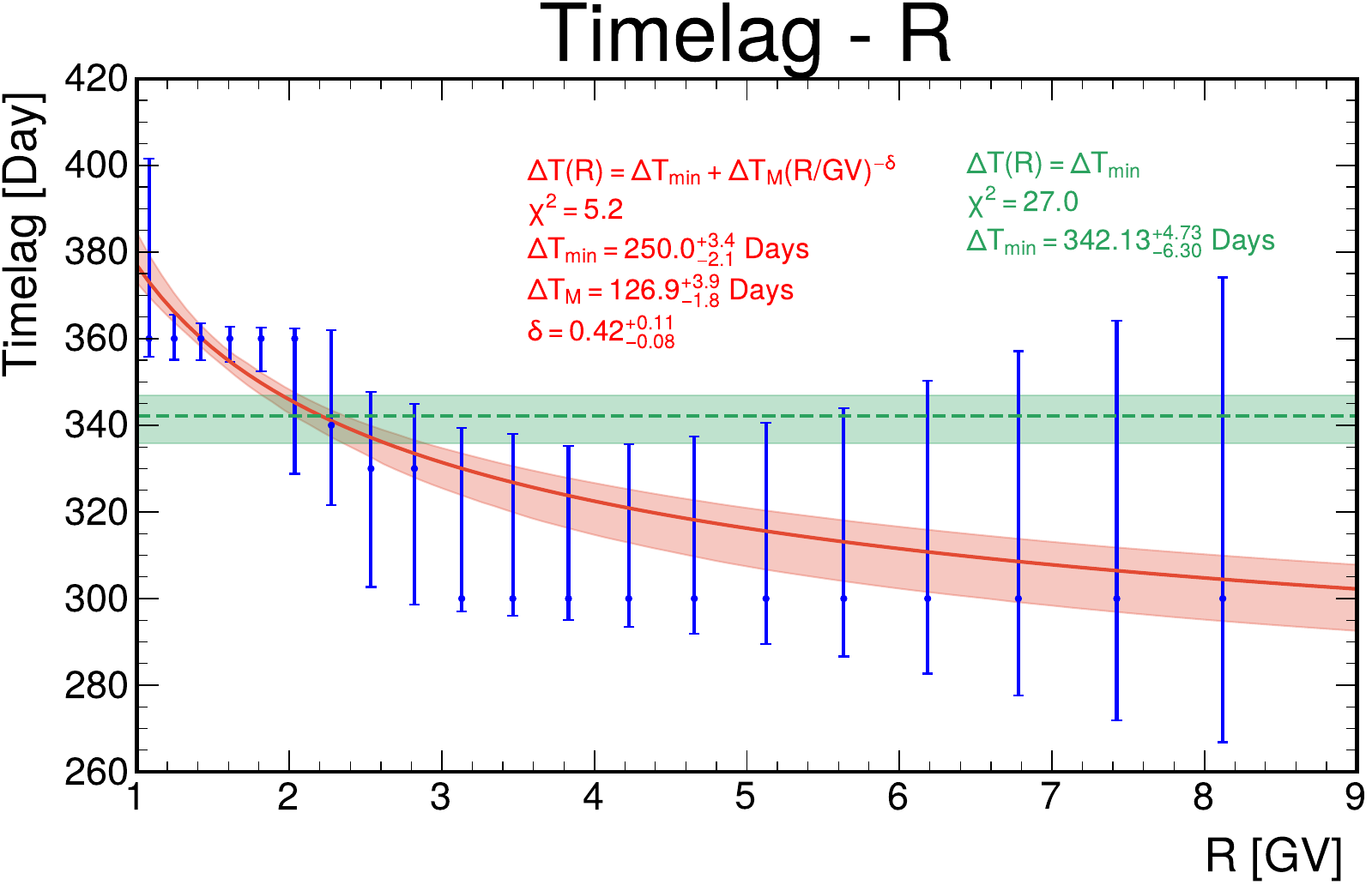}
\caption{
Fitted time lag data within the rigidity range 1-8.48 GV. The red curve represents the best-fit model for the time lag ${\Delta T}(R) = {\Delta T}_{\text{min}} + {\Delta T}_M (R/GV)^{-\delta}$, where ${\Delta T}_{\text{min}}$, ${\Delta T}_M$, and $\delta$ are fitted parameters. The blue points with error bars correspond to the measured time lag values and their uncertainties across different rigidity bins. Red curves represent the 68$\%$ confidence intervals of the R-dependent fit, while green horizontal lines indicate the 68$\%$ confidence intervals of the R-independent fit. }
\label{GV-Tfit}
\end{figure}
To further analyze the rigidity dependence of the time delay, we performed a power-law fit within a defined rigidity range using Eq.\ref{fit}. The fitting yielded a power-law index of 
$\delta = 0.42^{+0.11}_{-0.08},$ as shown in Fig.\ref{GV-Tfit}, suggesting a meaningful relationship. All fitting parameters are shown in Table.\ref{table1}. This result is consistent with similar findings reported in recent literature \citep{PhysRevD.106.103022}.

\begin{table}[htbp]
  \centering
  \caption{Comparison of Model Parameters}
  \setlength{\tabcolsep}{10pt}
  \label{tab:model_params}
  \begin{tabular}{ccc}
    \toprule
    \textbf{Parameters} & \textbf{R Dependent} & \textbf{R Independent} \\
    \midrule
    $\Delta T_{\text{min}} \text{[Day]}$  & $250.0^{+3.4}_{-2.1}$ & $342.13^{+4.73}_{-6.30}$ \\
    $\Delta T_M \text{[Day]}$            & $126.9^{+3.9}_{-1.8}$ & Null \\
    $\delta$                      & $0.42^{+0.11}_{-0.08}$ & Null \\
    \midrule
    $\chi^2$                      & 5.7                    & 27.0 \\
    \bottomrule
  \end{tabular}
\label{table1}
\end{table}

\begin{figure*}[t]
    \centering
    \includegraphics[width=0.8\linewidth]{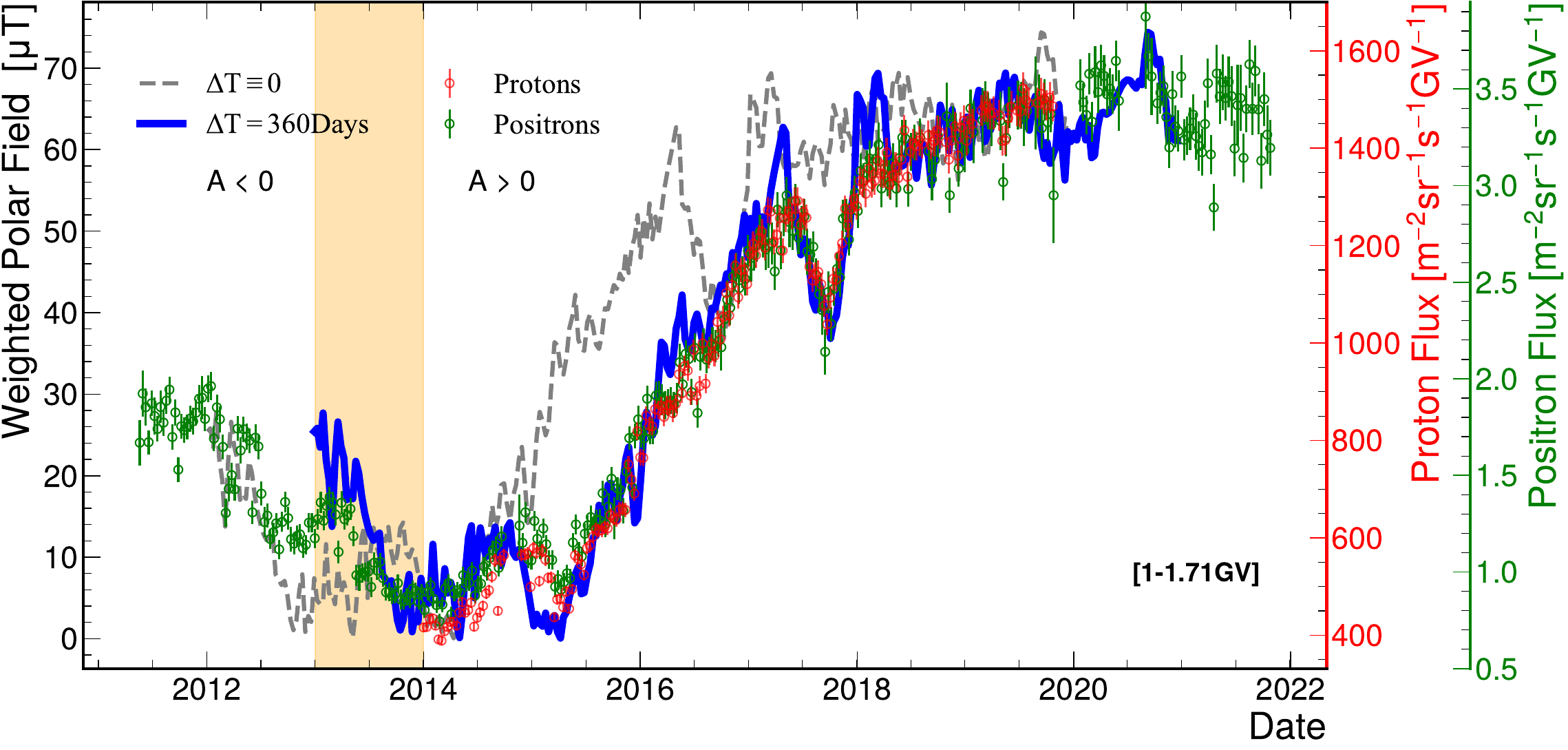}
    \caption{Time profile of the proton flux at Rigidity=1-1.71GV. Best-fit calculations are shown as a thick blue solid line, in comparison with the data \citep{PhysRevLett.127.271102}. Calculations for $\Delta T \equiv 0$ are shown as thin dashed lines. The orange shaded bars indicate the magnetic reversals of the Sun's polarity at 2013 \citep{2015ApJ...798..114S}.}
    \label{fig:your_label}
\end{figure*}

Fig.\ref{fig:your_label} shows the weighted polar field and the flux of galactic cosmic rays (protons and positrons) over time from 2011 to 2022. The fluxes of protons and positrons are displayed on the right axis in red and green, respectively. The shaded region marks the period of the polar field reversal, where the polarity change complicates the identification of a consistent time delay between galactic cosmic ray flux and magnetic field variations. From our study, we observed that during the 
$A>0$ period (2014-2020), there is a notable time delay in the response of galactic cosmic ray flux to changes in the polar field, which varies with rigidity \citep{PhysRevD.106.103022}. In contrast, during the $A<0$ period, no similar time delay was observed, suggesting different modulation effects under positive and negative polarity cycles.

For the $ A<0 $ period, since AMS data collection started in 2011, we can only discuss GCR flux variations from 2011 to 2014. It has been noticed that the modulation effect during the negative polarity period on negatively charged particles may be similar to the modulation effect on positively charged particles during the positive polarity period with the positron fraction data provided by AMS-01 and PAMELA \cite{DELLATORRE20121587}. 

\begin{figure}[h]
\centering
\includegraphics[width=1\linewidth]{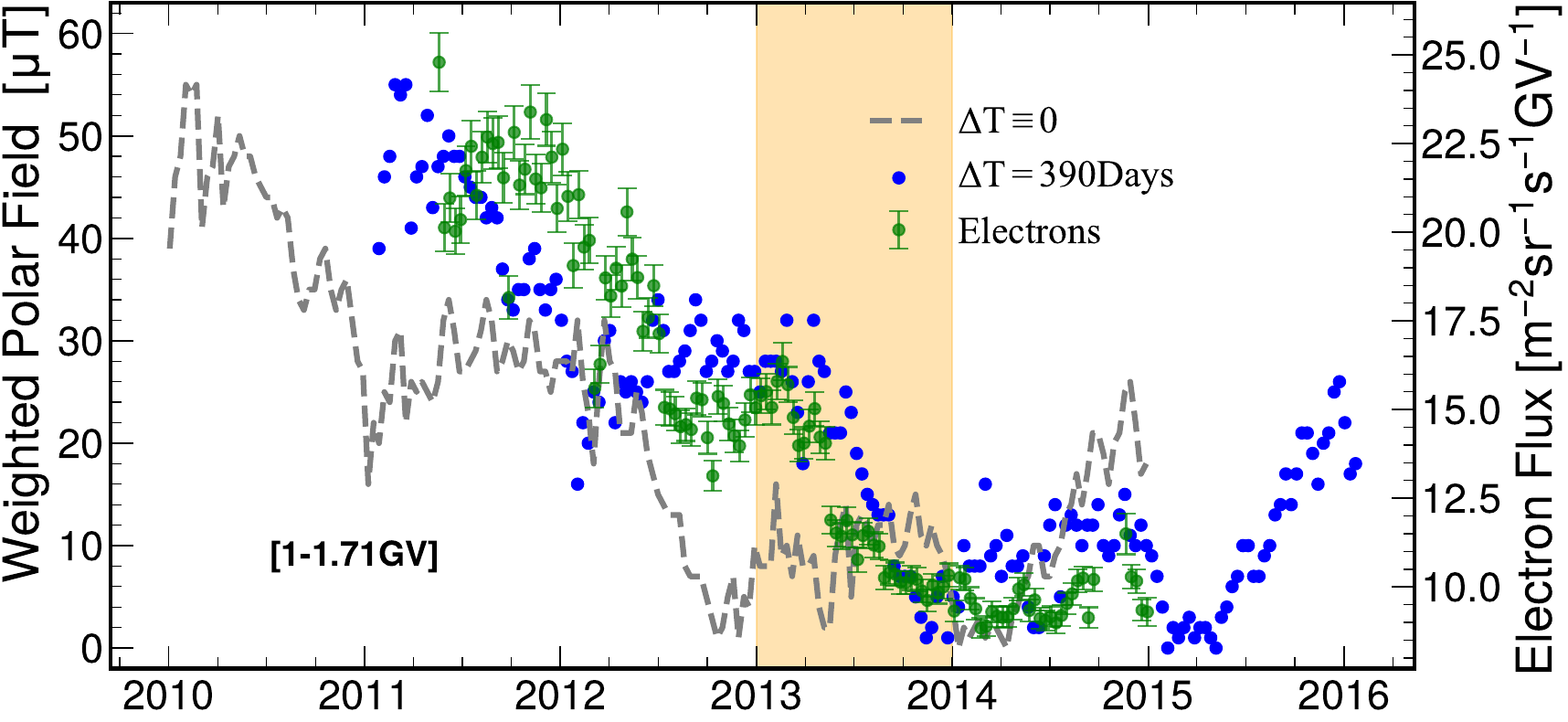}
\caption{Comparison of weighted polar field and electron flux (1-1.71 GV) over Time: Magnetic field data is shown with no time shift (grey) and with a time shift of 390 days (blue), compared to electron flux measurements (green) from 2011 to 2014. 
\label{elec}}
\end{figure}

Therefore, we also examined the delay relationship between electrons \citep{PhysRevLett.130.161001} and the polar field from 2011 to 2014, finding a similar time delay of $390^{+56}_{-7}$ days for electrons in the rigidity range of 1-1.71 GV as is shown in Fig.\ref{elec}. 

However, due to the later start of AMS data collection and the insufficient amount of data, we cannot provide a conclusive and accurate result. Future calculations can leverage the new data currently being measured by AMS \citep{AGUILAR20211} for the next solar cycle. Crucial tests can be performed by AMS through detailed measurements of individual particle fluxes for $p, \bar{p}, e^{+}, \text {and } e^{-}$. Such measurements, under varying polarity conditions and during polarity reversals, can offer deeper insights into the dynamics of GCR modulation and the effects of the polar field on galactic cosmic rays.

We have found that the polar field exhibits a delayed effect, and other literature suggests that the sunspot number also experiences a delay \citep{tomassetti2017evidence}. Additionally, various other solar parameters can be considered. In the future, we can apply machine learning methods to incorporate these different parameters to predict galactic cosmic ray behavior more effectively.
the N and S hemispheres to best describe the magnetic fields affecting galactic cosmic rays.

\section{Conclusion} \label{sec:conclusion}
In this letter, we find the correlations between the positive GCR fluxes and the polar field data from the North and South hemispheres. Our analysis primarily focuses on the 24th solar cycle post-2013, corresponding to the $A>0$ positive polarity period.  We defined a parameter $w=0.78$ to adjust the weights of the N and S hemispheres to best describe the magnetic fields affecting galactic cosmic rays. 

For the $A<0$ negative polarity period, the weight $w=1.05>1$ reflects the stronger influence of the southern hemisphere magnetic field on GCRs. This contrast in w values highlights the shift in magnetic field dominance between the $A>0$ and $A<0$ periods.

Previous studies \citep{Koldobskiy2022} have investigated the relationship between polar fields and Galactic cosmic-ray intensity, focusing on time lags among solar indices such as sunspot numbers (SSN), open solar flux (OSF), and cosmic-ray variations. These studies used cross-correlation and wavelet-coherence methods to identify coherence across solar cycles, particularly noting the 11-year and 22-year periodicities in the time lag influenced by heliospheric conditions. In contrast, our work specifically examines the 24th solar cycle post-2013, focusing on the impact of North and South hemisphere polar fields on GCRs. By using daily rather than monthly data, our analysis provides a finer time resolution, enhancing the precision of the observed correlations.

For the first time, we discovered a strong correlation between GCR fluxes and the combined North-South components of the WSO magnetic field. By defining a weighted parameter to adjust the influence of each solar hemisphere, 
our model accounts for asymmetries caused by the seasonal variations observed at Earth, which previous models did not fully capture.  This new parameter enables a more accurate representation of the time delay between Galactic cosmic rays and polar field variations, offering a more detailed view of modulation dynamics within the heliosphere.

 Our model successfully predicted the temporal evolution of cosmic-ray proton spectra in the 24th solar cycle after the polar field reversal as measured by the AMS experiment. Once the correlation between modulation parameters and solar activity indices is established, our model demonstrates high predictive accuracy. By utilizing extensive cosmic-ray proton data, our research uncovered a significant aspect of cosmic-ray modulation dynamics within the expanding heliosphere. Specifically, we identified a time lag $\Delta T$ of approximately 10 months between cosmic-ray data in the 1-33.5 GV range and polar field data, contingent on the heliospheric conditions.

Tomassetti, Bertucci, and Fiandrini \cite{PhysRevD.106.103022} have suggested that the time lag would be related to the solar wind. Here we propose a mechanism that would be able to explain the time lag qualitatively. The variation of Galactic cosmic ray strength is due to the modulation of the heliopause barrier for interstellar plasma flow.\cite{2021AGUFMSH22A..04R} The modulation is triggered by the variation of ions flux of solar wind that propagates to the heliopause which is about $L\sim 120$ AU from the sun. \cite{2019NatAs...3.1007B,2019NatAs...3.1019R,2019NatAs...3.1024G} It is known that solar wind speed $V\sim 600$ km/s. The propagation time of the ions flux, which contains the information of solar activity manifested by the observed magnetic field variation may be estimated as $\Delta T \sim L/V = 120\, AU/600\,km/s \sim 3.0 \times 10^7\,s \sim 350\, days $, which is consistent with the time lag found in \cite{tomassetti2017evidence, Jiang:2023cnx} and the present paper.

On the other hand, a GCR with 1 GV rigidity would take approximately one day to travel from the heliopause to Earth if its trajectory were straight, making the total time-lag of the same order as $\Delta T$. Although the trajectories of GCRs with small tilt angles relative to the heliospheric current sheet deviate significantly from a straight path, which is known as the drift effects, those with large tilt angles can be approximated as straight since magnetic effects are negligible. However, we have not yet considered the diffusion process, which plays an important role when particles undergo multiple scatterings with ions in the heliosphere.

In our model, the galactic cosmic ray propagates through the heliosphere directly, following, for instance, the interstellar magnetic field. As a result, drift and diffusion effects are negligible. We should also highlight another model proposed in \cite{PhysRevD.106.103022}, which associates the time lag with the diffusion of GCR in the magnetic field inside the heliosphere and shows that the diffusion coefficient is rigidity-dependent. We also perform an analysis on the rigidity dependence of time lag with their model and obtain the coefficient of $\delta =0.42^{+0.11}_{-0.08}$. It does not oppose the phenomenological model of \cite{PhysRevD.106.103022} but the uncertainty of our data is too large to make any conclusion. We suggest that this phenomenon would be due to the rigidity dependence of the heliopause barrier, at least partly.
% oting that the \textcolor{red}{\marginpar{A4}galactic} cosmic ray of \sout{rigidity larger than 1 GV takes only a few minutes} \textcolor{red}{1 GV would take about 1 day} in the journey from the heliopause to the earth \textcolor{red}{if we only consider the straight trajectories, thought there would be trajectories very different from straight lines}.

A noteworthy outcome of our findings is the ability to predict the galactic cosmic-ray flux at Earth using solar activity indices observed at the time $t-\Delta T$. This capability is crucial for real-time space weather forecasting, an important consideration for human spaceflight.

In this work, the parameter $\Delta T$  was determined by correlating AMS measurements of cosmic-ray protons from 2014 to 2019 with polar field data sampled every 10 days. Future research could include the use of neutron monitor data \citep{Smith2022, MAURIN2015363}, longer observation periods, or accounting for periodicities and latitudinal dependencies in polar field measurements \citep{POTGIETER1997883}. Due to the 10-day aggregation and absence of error estimates in the polar field data, we could not test these additional hypotheses. Using daily data and including measurement errors would significantly enhance the analysis.
However, a detailed re-analysis of our model will be possible with the forthcoming monthly resolved data from AMS on cosmic-ray particle and antiparticle fluxes.

The results of this study provide significant insights into the dynamics of cosmic-ray modulation and offer practical implications for space weather prediction, enhancing the safety and planning of human space missions.

\begin{acknowledgments}
We are grateful for important physics discussions with Zhi-Cheng Tang and Su-Jie Lin. This work is supported in part by the Guangdong Provincial Key Laboratory of Advanced Particle Detection Technology (2024B1212010005), the Guangdong Provincial Key Laboratory of Gamma-Gamma Collider and Its Comprehensive Applications (2024KSYS001), the Fundamental Research Funds for the Central Universities, and the Sun Yat-sen University Science Foundation.
\end{acknowledgments}

\bibliography{apssamp}{}% Produces the bibliography via BibTeX.
\bibliographystyle{apsrev4-2}

\end{document}